# Superconductivity in high-pressure synthesized pure and doped MgB$_2$ compounds

**P. Toulemonde,**

N. Musolino, H.L. Suo and R. Flükiger.

*Département de Physique de la Matière Condensée, Université de Genève,*

*24 quai Ernest-Ansermat, CH-1211 Genève 4, Switzerland.*

**Abstract**

Dense pure and doped $(Mg_{1-x}A_x)B_2$ samples with A = Na, Ca, Cu, Ag, Zn and Al were synthesized at high pressure – high temperature in a multi anvils press (3.5-6 GPa, 900-1000°C) for $0 < x ? 0.20$. They were studied by x-ray diffraction, scanning electron microscopy and their superconducting properties were investigated by a.c. susceptibility, magnetization and transport measurements. Only Al is really substituted on the Mg site. The others elements form secondary phases with B or Mg. No large effect is observed on the superconducting properties $T_c$, $j_c$ critical current, $H_{irr}$ and $H_{c2}$.

**Introduction**

Since the discovery of superconductivity in $MgB_2$ at 39 K [1], an intense effort has been focused on the preparation of high quality samples. Due to the low thermal stability of $MgB_2$, its preparation has to be done in closed systems [2, 3]. The method is based on the reaction between a fine powder of boron and liquid magnesium in equilibrium with its vapor above the melting point of Mg (650 °C) in a sealed metallic tube of Ta or Mo. This technique at low pressure is currently used to prepare $MgB_2$ pellets [4] (or wires by diffusion of Mg into B fibers [5]), but the polycrystalline samples obtained are porous and not suitable for good transport measurements. Using a high pressure process, it is possible to synthesize dense $MgB_2$ samples. For over 40 years, magnesium has been used as a catalyst for the transformation of hexagonal boron nitride (BN) to its cubic form at high pressure – high temperature (HP-HT) [6-8] ; during this HP-HT treatment the formation of secondary $MgB_6$ and $MgB_2$ phases was observed [9]. Recently, the HP-HT process has been applied to $MgB_2$ for its sintering [10-12], its synthesis [10, 13], and its single crystal growth [14-16]. In this paper the HP-HT preparation and the superconducting properties of pure $MgB_2$ and doped $(Mg_{1-x}A_x)B_2$ with A = Al, Zn, Cu, Ag, Na, Ca are discussed.

**1. Pure $MgB_2$ compounds**

**1.1. High pressure sintering and HP-HT synthesis**

As other groups [11-13], we have used the HP-HT process to sinter dense and high quality $MgB_2$ samples, starting from a commercial $MgB_2$ powder (Alfa, 98 %). Our samples were suitable for critical current measurements by magnetization and transport techniques. One year ago, we showed that the critical current in dense polycrystalline $MgB_2$ samples is not sensitive to the grain boundaries [10]. Moreover, shortly after the discovery of superconductivity in $MgB_2$, we were among the first to synthesize $MgB_2$ at HP-HT, starting directly from metallic Mg and B powders [10]. The method consists of a reaction at HP-HT of Mg and B. The powders were mixed together in a stoichiometric ratio, inserted into a BN crucible and placed in a cubic high pressure cell made of pyrophyllite. A cubic multi-anvils press was used to apply a pressure of 3-4 GPa and the temperature was maintained at 900-1000 °C for 1-2 h during the HP-HT synthesis. The conditions (pressure, time and temperature) were optimized to obtain the sharpest superconducting transition.

**1.2. Characterization**

The samples were characterized by x-ray diffraction using copper $K_{\alpha}$ radiation ($\lambda$ = 1.544 Å) ; the XRD patterns of three different samples are presented in figure 1. These samples were prepared with similar P, T conditions (3.5-4.5 GPa, 850-950 °C) from Mg and B of different granulometry : flakes or fine powder (Alfa, 99.8%, 325mesh) for Mg and crystalline powder for B (Ventron, m2N7, 60 mesh and Alfa, 99.7%, 325 mesh). The main impurity is MgO, whose concentration decreases for finer size starting Mg powders. Combining crystalline fine Mg and B powders gives the purest sample. This MgO impurity is present also in the doped samples (paragraph 2.2.) but its proportion is small compared to the other impurities based on the doping element (figure 4). The lattice parameters of the HP-HT synthesized $MgB_2$ samples, calculated by a least square method from the reflections list, are consistent with the literature [17] : a = 3.083(4) Å, c = 3.518(2) Å. The microstructure of a typical sample is shown figure 2 ; it consists of small, well shaped $MgB_2$ grains of 1-10 μm and small MgO particles. Back scattered electron images show evidence of boron rich dark regions (around 10 μm) corresponding to not yet fully reacted boron grains. These regions explain why the apparent $j_c$ value is lower for synthesized samples than for sintered samples from a commercial $MgB_2$ powder (higher than $10^6$ A/cm$^2$ at 4.2 K, in zero field) ; indeed, the whole volume of the sample (including the non superconducting boron rich regions) is taken into account in the $j_c$ calculations from the magnetization data obtained in a VSM and analyzed using the Bean model.

The figure 3 shows the a.c. susceptibility of the three samples. The best results are obtained for the mixture of very fine reacting powders (325 mesh ~ 40 μm). The transition is very broad (around 10 K) for the sample prepared from Mg flakes and its width decreases as the size of the reacting powders decreases, reaching ~ 1.5 K for the best sample. The $j_c$ critical current measured for this sample by magnetization (VSM) was estimated to be 2?$10^5$ A/cm2 at 4.2 K in zero field and $10^4$ A/cm$^2$ at 20 K, in a magnetic field of 2 T (see figure 8 in ref. 10).

## 2. Doping on Mg site in MgB$_2$

Initially, different substitutions on the Mg site were tested experimentally [18-29] and theoretically [30-32] in attempt to increase the $T_c$ of MgB$_2$. Additive elements were also investigated to increase the pinning potential of the material in order to reach higher $j_c$ values. Except for aluminum [18], few elements enter the MgB$_2$ lattice and substitute on the Mg site to give a real solid solution (Mg$_{1-x}$A$_x$)B$_2$.

### 2.1. HP-HT synthesis

The ambient pressure used in most syntheses remains quite low and could limit the possibilities of substitutions. The HP-HT process may increase the solubility limit of the element in substitution for Mg. The different substitutions tried were A = Al, Zn, Cu, Ag, Na, Ca with x = 0.1 in our multi-anvils press. Other compositions for 0 < x < 0.20 were studied for Al and Zn. The additive element was mixed with the Mg and B fine powders in a stoichiometric ratio and the synthesis conditions were kept close to those for pure MgB$_2$ samples (table 1).

## 2.2. Characterization

Figure 4 shows the XRD patterns of MgB$_2$ doped samples with x = 0.10 of Na, Ca, Cu, Zn and Ag. The main phase is still MgB$_2$, but some impurities appear : NaCl, CaB$_6$, Cu$_2$Mg, Zn + MgZn$_2$ and MgAg respectively. The lattice parameters of the MgB$_2$ phase in these doped samples do not change significantly and stay very close to the initial values for undoped materials (table 1.). This means that these elements do not substitute for Mg. In particular for Zn, considering the ionic radius of Mg (0.65 Å) and Zn (0.74 Å), one could expect a continuous decrease of the c-axis in the Zn-doped series, around 0.02 Å for 10 % of substitution. However, and contrary to Kazakov et al. [19], we found no evidence for a change in the a and c lattice parameters in our high-pressure prepared samples, within experimental errors. If Zn really occupies the Mg site, its solubility limit is low.

SEM images (figure 5) confirm the presence of the secondary phase detected by XRD. The typical size of these impurity grains varies from 1 (MgO) to 20 µm (for Cu$_2$Mg or MgZn$_2$ for instance).

A.C. susceptibility measurements show that these doped samples have a slightly lower T$_c$, possibly due to under-stoichiometry on Mg site. The samples show a continuous broadening of the superconducting transition with increasing concentration of Zn and Al (table 1).

In Al doped (Mg$_{1-x}$Al$_x$)B$_2$ materials, the lattice parameters decrease, as expected [18] : 0.21 Å/x for a-axis and four times faster for c-axis : 0.88 Å/x. A gradual decrease of T$_{c\,onset}$ from 38 K (x = 0.01) to 36 K (x = 0.10) with increasing Al content is also measured below x = 0.10 (in the single-phase region [18]), but the transitions (from SQUID measurements, not shown) do not broaden as much as in ref. [18] (table 1). This suggests that the substitution of Al on the Mg site is more homogeneous in our samples.

The defects made up of the grains of secondary phases (for Na, Ca, Cu, Zn and Ag) could act as pinning centers and may increase the critical current. However, the magnetization measurements (VSM) show that these inclusions do not really improve $j_c$ and its magnetic field dependence. The curves displayed in figure 6 for 5 % Zn doped, 10 % Na doped and pure $MgB_2$ are close to each other. The size of the impurity grains (> 1 µm) is greater than the typical coherence length of $MgB_2$, so they can not be efficient pinning centers. In addition, when the solid solution really exists, i.e. for Mg site occupied by Al, the $j_c$ behavior as a function of magnetic field is degraded (see curve of 5 % Al doped sample, figure 6).

In the case of Al, where a continuous solid solution exist up to x = 0.10, the local defects introduced by Al on the Mg site should decrease the mean free path "l" of the charge carriers and thus increase the resitivity ? (in the normal state), and perhaps increase $H_{c2}$ which is proportional to $?_0/(?_0 l)$. For x = 0.05, resistivity measurements were performed under different magnetic fields up to 14 T and $H_{c2}$ was extracted from the onset of the superconducting transition and the irreversibility field from the offset. They were compared to $H_{c2}$ and $H_{irr}$ determined from the shift in $T_c$ in the Meissner regime measured under different fields and magnetization loops in a SQUID magnetometer. No significative increase of $H_{irr}$ is observed (extrapolated value : $H_{irr}$ (0K) ~ 10 T), compared to the value for pure $MgB_2$ [10], but a slight increase of extrapolated $H_{c2}$ (0 K) from ~ 18 T (x = 0, [10]) to ~ 21 T for x = 0.05 was observed (figure 7.). This result has to be confirmed for x ? 0.05.

3. **Conclusion**

Sintered and synthesized pure $MgB_2$ and doped $(Mg_{1-x}A_x)B_2$ compounds (A = Na, Ca, Cu, Zn, Ag and Al for x ? 0.20) were prepared at high temperature - high pressure using a multi-anvils press. This HP-HT process allows us to obtain dense polycrystalline samples suitable for transport measurements. XRD and SEM investigations show that only Al is substituted for Mg. The other doping elements react with Mg or B to form secondary phases. The $T_c$ value is largely unaltered, except in the case of Al where it decreases slightly with the doping level. The largest effect is a broadening of the superconducting transition. The magnetization measurements show that $j_c$ is not improved and $H_{c2}$ only slightly increased for the Al substituted samples. The impurity inclusions and Al distribution on the Mg site are not efficient pinning centers. This work suggests that the origin of increased critical currents reported in bulk samples is perhaps due to defect pinning centers (oxygen, carbon) in the boron plane. This is an area of current research.

**Acknowledgment**

P. Toulemonde would like to thank A. Naula for his help in the technical preparation of the HP-HT experiments performed in the multi-anvils press.

**Figure captions**

**Figure 1.** X-ray diffraction pattern of high pressure – high temperature synthesized $MgB_2$ samples with Mg and B of different granulometry at ? = 1.544 Å.

**Figure 2.** SEM image of HP-HT synthesized $MgB_2$ with crystalline fine Mg and B powders (325mesh).

**Figure 3.** A.C. susceptibility measurements of the superconducting transition for the HP-HT synthesized $MgB_2$ samples.

**Figure 4.** X-ray diffraction pattern at ? = 1.5418 Å of HP – HT synthesized $(Mg_{1-x}A_x)B_2$ samples with A = Na, Ca, Cu, Zn and Ag, x = 0.10.

**Figure 5.** Electron back scattered image of HP-HT synthesized $(Mg_{0.90}Zn_{0.10})B_2$ sample.

**Figure 6.** Inductively measured $j_c$ values plotted against applied magnetic field for different temperature (10, 20 and 30 K) of 5 % Zn, 5 %Al, 10 % Na doped and pure $MgB_2$ samples.

**Figure 7.** The temperature dependence of the second critical field $H_{c2}$ and of the irreversibility field $H_{irr}$ measured inductively (VSM or SQUID data) for $(Mg_{0.95}Al_{0.05})B_2$ sample.

**Tables**

**Table 1.** HP-HT synthesis conditions, inductive $T_c$, and lattice parameters of the $(Mg_{1-x}A_x)B_2$ samples with A = Na, Al, Ca, Cu, Zn and Ag for 0 < x ? 0.20.

| doping element | content x | synthesis conditions | $T_c$ onset (K) | $\Delta T_c$ (K) | lattice parameters a (Å) | c (Å) |
|---|---|---|---|---|---|---|
| **Na** (NaCl) | 0.10 | 4 GPa, 950°C, 1h | 38 | 3 | 3.082(5) | 3.517(8) |
| **Al** | 0.01 | 3.5 GPa, 950°C, 1h | 38 | 3.5 | 3.084(1) | 3.517(1) |
| | 0.03 | 3.5 GPa, 950°C, 1h | 38 | 3.5 | 3.081(1) | 3.511(1) |
| | 0.05 | 3.5 GPa, 950°C, 1h | 37 | 4.5 | 3.078(2) | 3.492(2) |
| | 0.10 | 3.5 GPa, 900°C, 1h | 36 | 6 | 3.065(5) | 3.437(2) |
| **Ca** (CaH$_2$) | 0.10 | 3.5 GPa, 900°C, 1h | 38 | 3 | 3.084(2) | 3.526(1) |
| **Cu** | 0.10 | 4 GPa, 900°C, 1h | 38.5 | 2 | 3.076(4) | 3.519(3) |
| **Zn** | 0.05 | 3.5 GPa, 950°C, 1h | 39 | 4 | 3.085(2) | 3.523(2) |
| | 0.10 | 3.5 GPa, 950°C, 1h | 38 | 2.5 | 3.083(1) | 3.523(1) |
| | 0.20 | 6 GPa, 900°C, 1h | 38 | 4 | 3.082(8) | 3.513(7) |
| **Ag** | 0.10 | 3.5 GPa, 1000°C, 1h | 38.5 | 2.5 | 3.080(1) | 3.518(1) |

Table 1.

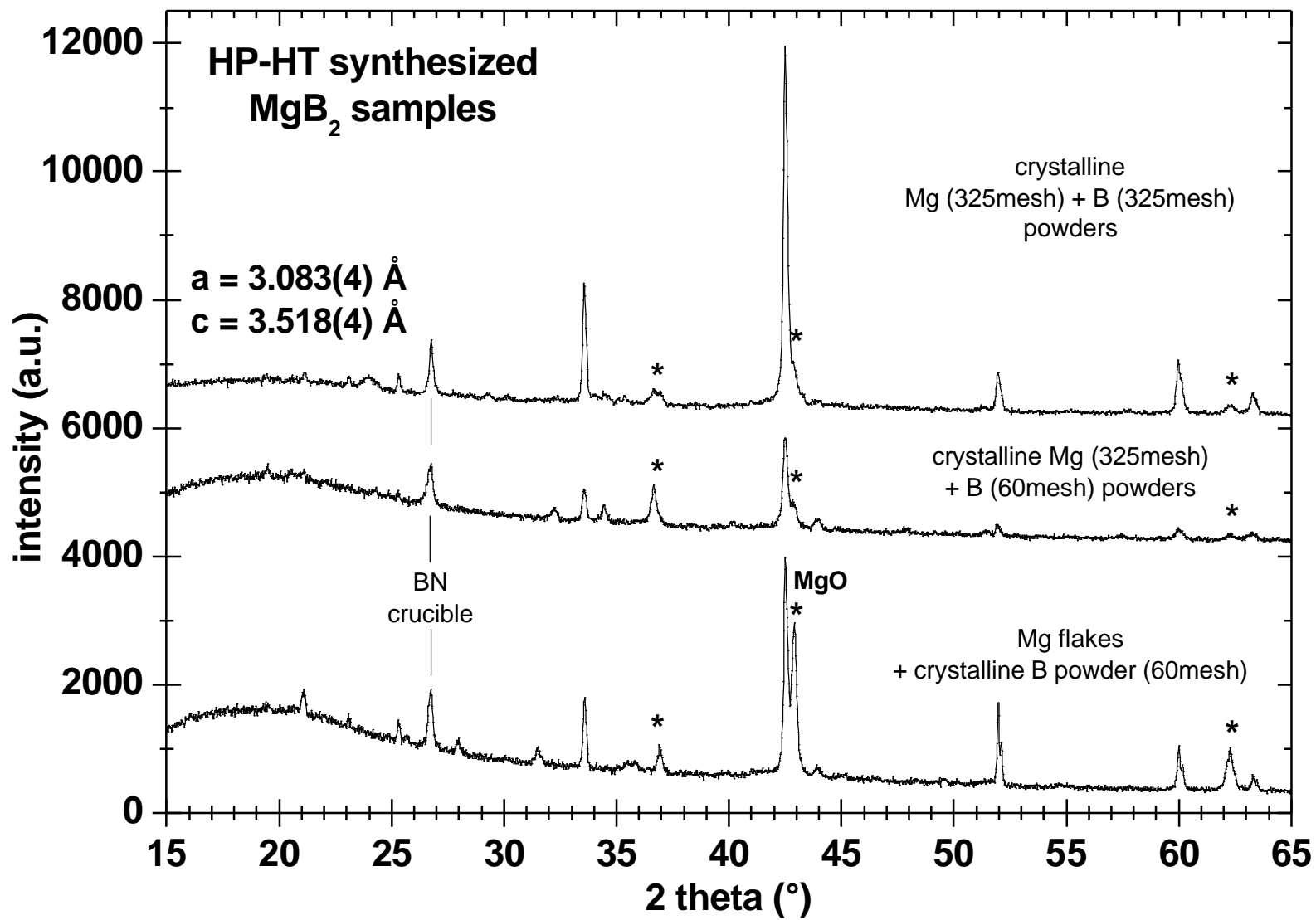

Fig. 1.

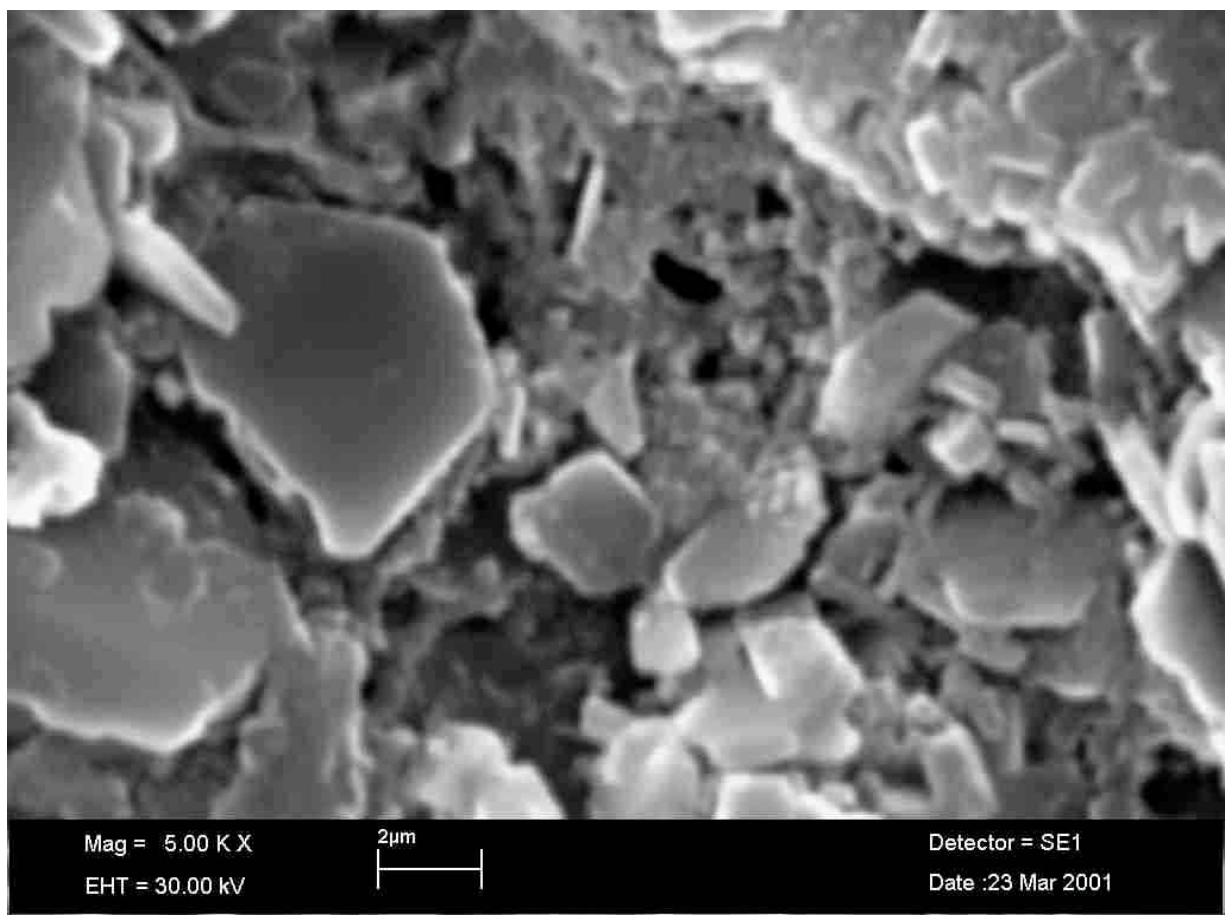

Fig. 2.

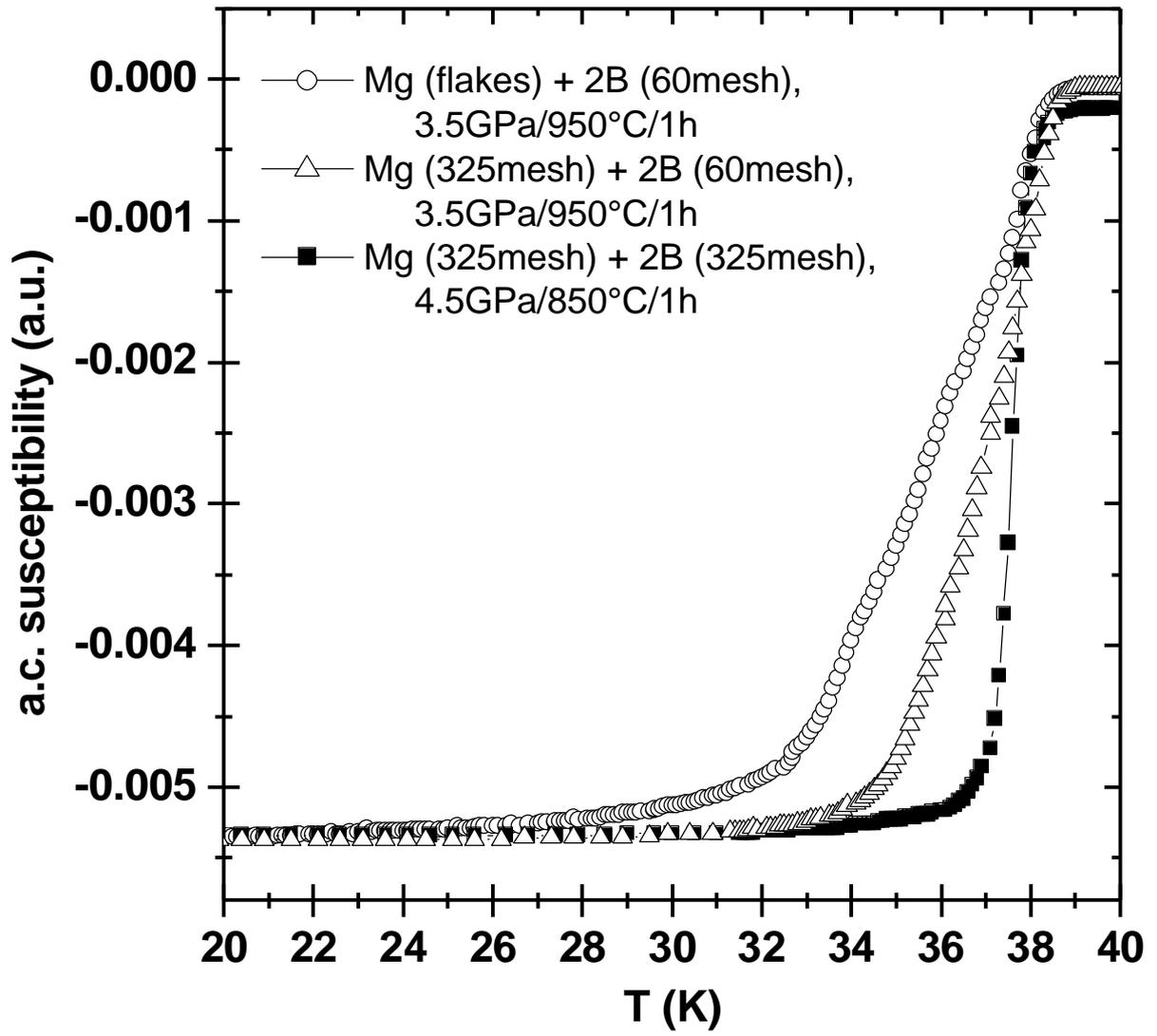

Fig. 3.

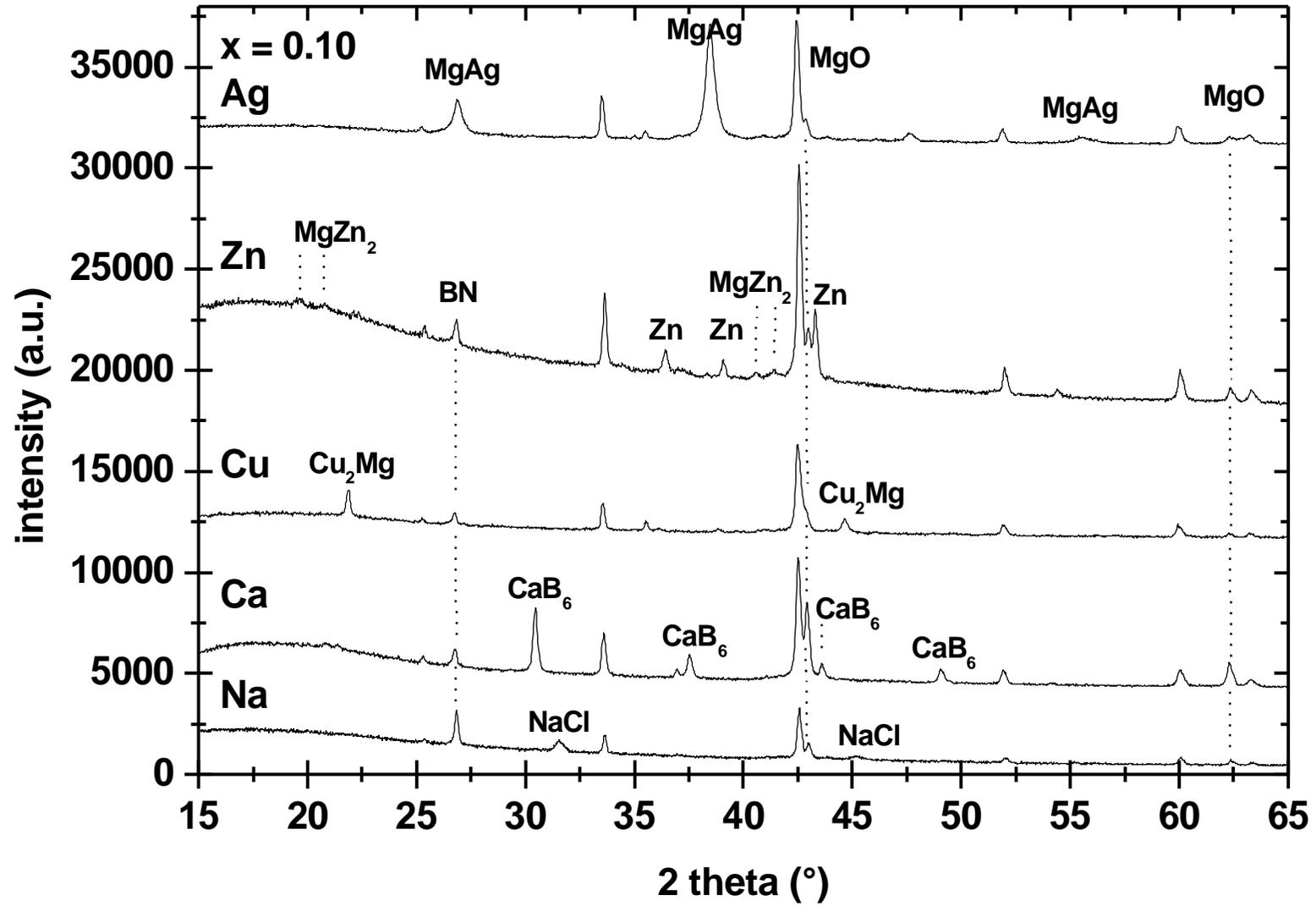

Fig. 4.

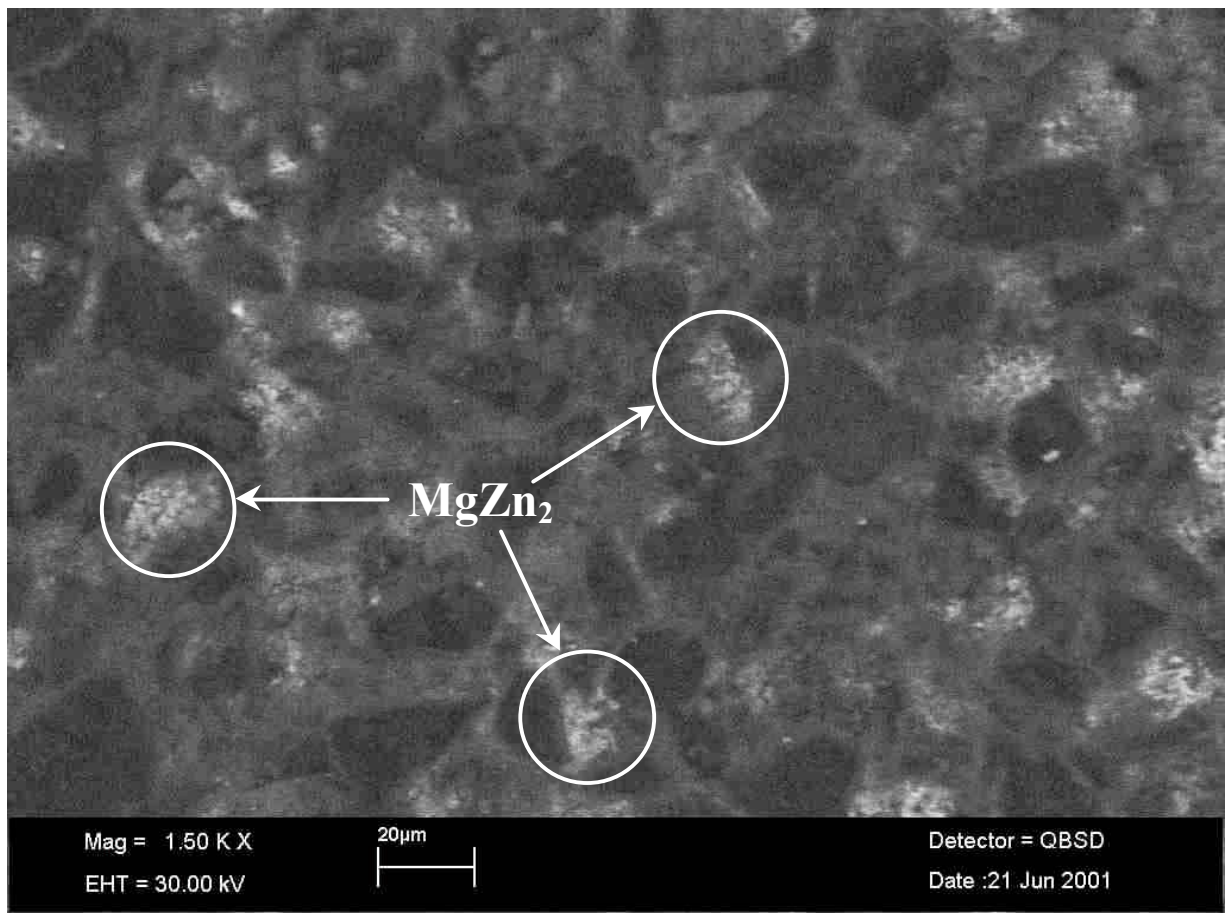

Fig. 5.

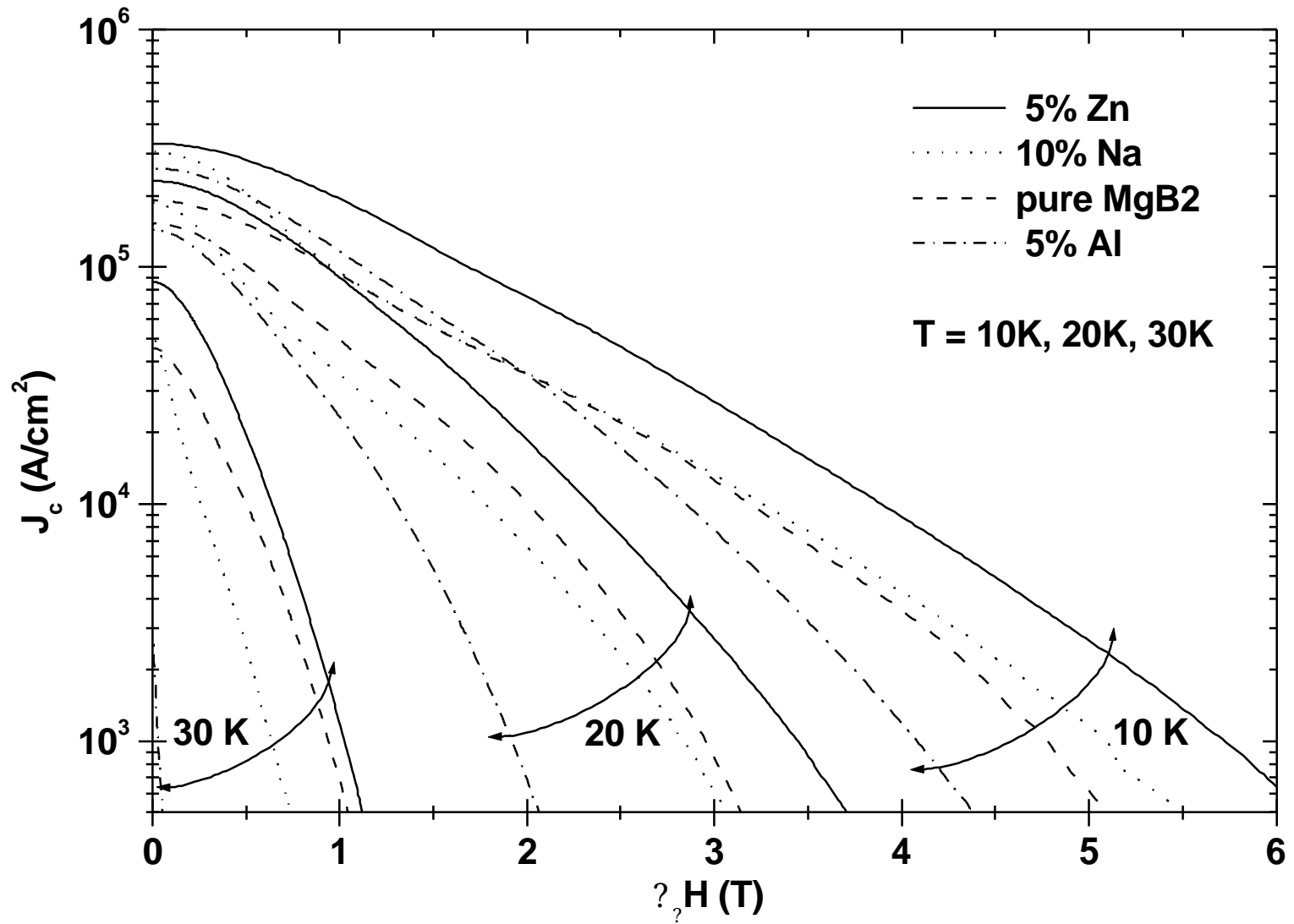

Fig. 6.

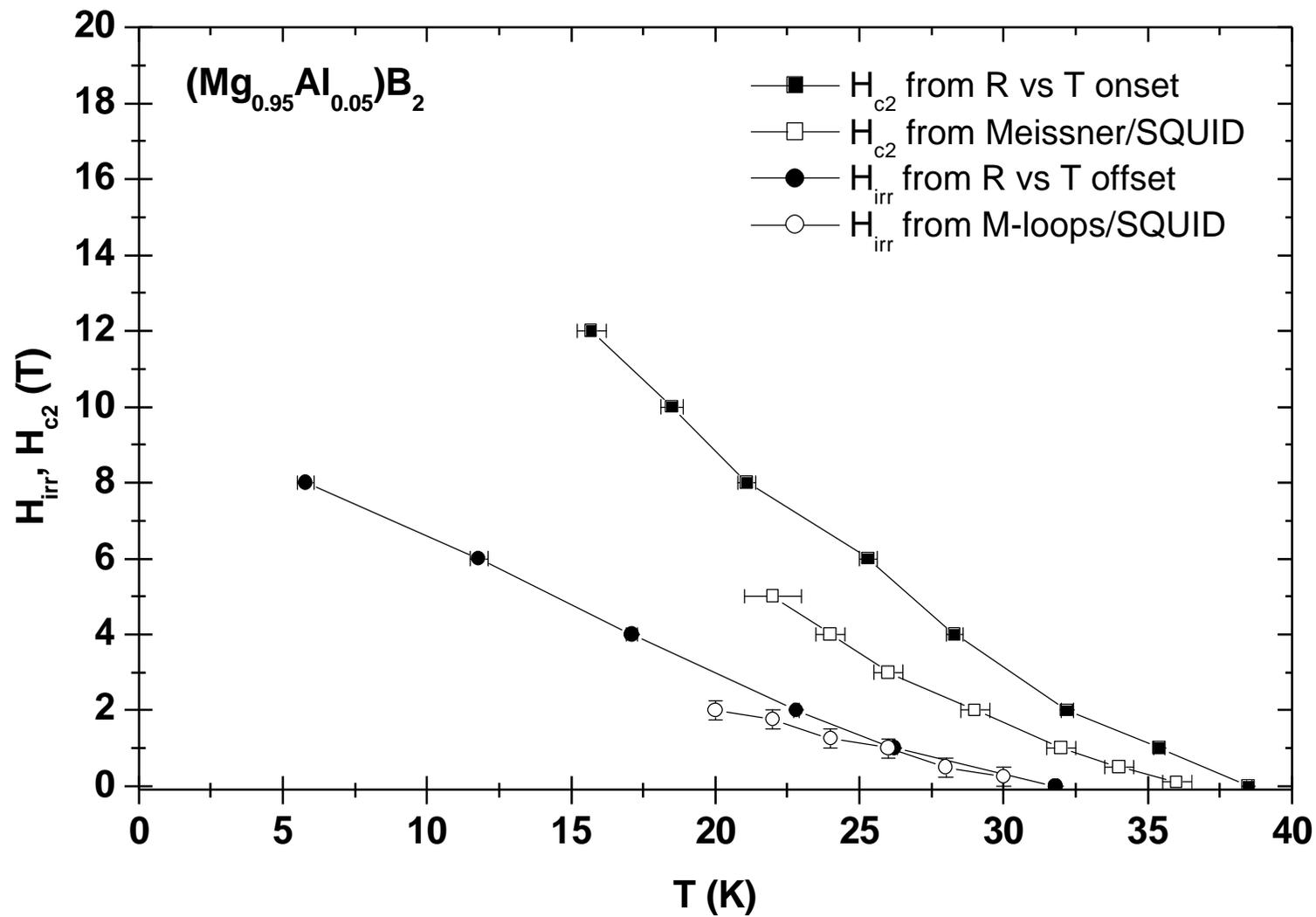

Fig. 7.